# Calculation of phases of *np-* scattering for potentials Reid93 and Argonne v18 on the phase-function method

V. I. Zhaba

Uzhgorod National University, Department of Theoretical Physics,
54, Voloshyna St., Uzhgorod, UA-88000, Ukraine

**Abstract:**

A known phase-functions method has been considered for calculation of a single-channel nucleon-nucleon scattering. The phase shifts of *np-* scattering turn out for modern realistic phenomenological nucleon-nucleon potentials Reid93 but Argonne v18. Influence of choice of numeral method is explored on the decision of phase equation. For the decision of phase equation as differential equation of 1th order numeral methods are used: Euler method and Runge-Kutta methods 2-, 3-, 4- and 5th orders of exactness. Numerically calculated phase shifts of *np-* scattering for potentials Reid93 and Argonne v18 are in good agreement with the results obtained with the other methods of original works. Based on the known phases of scattering one can obtain the full amplitude, the full cross-section and the partial scattering amplitude accordingly. The used Runge-Kutta methods can be applied for finding of phases shifts of the mixed channels of nucleon-nucleon scattering.

## 1. Introduction

Based on the experimentally observed values of the scattering cross-section and energies of transitions we get information about the scattering phases and amplitudes in the first place, than about the wave functions, which are the main object of the research in a standard approach. In other words, not the very wave functions are being observed in the experiment, but their changes caused by the interaction [1, 2]. It is therefore of interest to obtain the equations directly connecting the phases and scattering amplitudes with the potential without finding the wave functions.

The precise solution of the scattering problem aiming at calculation of the scattering phase is possible only for individual phenomenological potentials. When realistic potentials are used, the phases of scattering are roughly calculated. This is due to the use of physical approximations or numerical calculation. The impact of the numerical algorithm choice on the solution of scattering task is specified in Ref. [3]. The methods of solving the Schrödinger equation with the aim of obtaining the scattering phases include: the method of successive approximations, the Born approximation, the Brysk's approximation, the phase-functions method, and others.

The calculation results of the phase shifts of a single-channel nucleon-nucleon scattering for potentials Nijmegen group (Nijm I, Nijm II and Nijm 93) are given in Ref. [4]. The Runge-Kutta method of the fourth order was chosen as the numerical method of solving the phase equation.

This paper deals with the calculation of the phase shifts and full cross-sections of the nucleon-nucleon scattering for the up-to-date realistic phenomenological nucleon-nucleon Nijmegen potentials Reid93 and Argonne v18 by using the phase-functions method (PFM). We investigate the impact of the numerical algorithm choice on the solution of the phase equation.

## 2. The phase-functions method

Let us consider the scattering of a spin-free particle with the determined values of energy $E$ and the orbital moment $l$ in a spherical-symmetric potential $V(r)$. The Schrödinger equation for the corresponding radial wave function $u_l(r)$ takes the form [1]:

$$u''_l(r) + \left(k^2 - \frac{l(l+1)}{r^2} - U(r)\right) u_l(r) = 0, \qquad (1)$$

where $U(r) = 2mV(r)/\hbar^2$ – the renormalized interaction potential, $m$ – the reduced mass, $k^2 = 2mE/\hbar^2$ – the wave number.

PFM is a special method to solve the radial Schrödinger equation (1), which is a second order linear differential equation. It is quite convenient for obtaining scattering phases, because this method does not require calculating radial wave functions of scattering problem in a wide range firstly and then finding these phases by their asymptotics [1, 2].

Two linearly independent solutions of the free Schrödinger equation (1) (given $U\equiv 0$) are the known Riccati-Bessel functions $j_l(kr)$ and $n_l(kr)$. Only the regular solution $j_l(kr)$ at the point $r=0$ corresponds to the free motion, and in this case, asymptotically given the large values of $r$, the solution takes the form

$$u_l(r) \approx const \cdot \sin(kr - l\pi/2).$$

The presence of the potential leads to the fact that now in the area where the potential $U(r)$ disappears, the wave function includes a supplement of irregular solution of the free equation $n_l(kr)$. The degree of this supplement, which quantitatively describes the effect of interaction, is the very scattering phase $\delta_l$ [1];

$$u_l(r) \to const \cdot \sin(kr - l\pi/2 + \delta_l), \; r \to \infty.$$

The standard method of calculating the scattering phases is a solution of the Schrödinger equation (1) with the asymptotic boundary condition. PFM is the transition from Schrödinger equation to the equation for the phase function. For that purpose one should change [1, 2]:

$$u_l(r) = A_l(r)\left[\cos\delta_l(r) \cdot j_l(kr) - \sin\delta_l(r) \cdot n_l(kr)\right]. \qquad (2)$$

The two new introduced functions $\delta_l(r)$ and $A_l(r)$ are the corresponding scattering phases and normalization constants (amplitudes) of wave functions for scattering on a determined sequence of truncated potentials. $\delta_l(r)$ and $A_l(r)$ are called a phase and an amplitude function according to their physical content. Equations for phase function with the initial conditions are [1]:

$$\delta'_l = -\frac{1}{k}U(r)\left[\cos\delta_l(r) \cdot j_l(kr) - \sin\delta_l(r) \cdot n_l(kr)\right]^2, \quad \delta_l(0) = 0; \qquad (3)$$

The phase equation (3) was obtained for the first time by Drukarev [5], and then independently in the works of Bergmann, Kynch, Olsson, and Calogero.

## 3. Potentials of nucleon-nucleon interaction

Why it has been chosen to calculate the phase shifts using Nijmegen group potentials (NijmI, NijmII, Nijm93, Reid93 [6]) and Argonne v18 [7] potential? The parameters of potential models have been optimized in such a way that the value of $\chi^2$ in the direct fit to the data was minimized. The first improvement of the Nijm78 potential [8] began in the early 1990s. An improved version of the potential was called Nijm92pp, since it was updated for 1787 *pp* data. For potential Nijm92pp the value of $\chi^2/N_{pp}$ was 1.4. The following improvement of Nijm78 potential for np data gave Nijm93 model: $\chi^2/N_{pp}$ =1.8 for pp 1787 and $\chi^2/N_{nn}$=1.9 for 2514 *np* data, i.e. $\chi^2/N_{data}$=1.87. For the Nijm I and NijmII potentials the value of $\chi^2/N_{data}$=1.03. The original potential Reid68 [9] was parameterized on the basis of the phase Nijmegen group analysis and was called Reid93. Parameterization was conducted for 50 parameters $A_{ij}$ and $B_{ij}$ of the potential, with $\chi^2/N_{data}$=1.03 [6]. The structure and recording of the Reid93 potential is quite cumbersome.

The Argonne v18 potential [7] with 40 controlled parameters gives the value of $\chi^2/N_{data}$=1/09 for 4301 *pp* and *np* data in the energy range of 0-350 MeV. For the CD-Bonn potential [10] the value $\chi^2/N_{data}$ makes 1.01 for 2932 *pp* data and 1.02 for 3058 *np* data. Such potentials as Hamada-Johnston-62, Yale group potentials, Reid68, UrbanaV14, etc. have greater values $\chi^2$, since parameterized on the basis of narrower energy interval.

Therefore, Reid93 and Argonne v18 potentials are one of those realistic phenomenological potentials, which best describe nucleon-nucleon interaction. When calculating phases of scattering, one should consider the peculiarities of potentials of internucleonic interaction.

## 4. Numerical methods for solving of the phase equation

The phase equation (3) is a first order nonlinear differential equation. The numerical methods for solving differential equations of the 1st order include: Euler method, modified Euler method, Euler method with recalculation (method Gyun), Runge-Kutta methods, multistep methods (e.g. Adams method of the 4th order) [11-15]. These modifications of the Euler method belong to the group of methods for prediction and correction. Euler method is a one-step method of 1st order accuracy. Error of the method Gyun is proportional to the square of the integration step $h^2$. Runge-Kutta algorithms of the 3rd and 4th order with order errors respectively $h^3$ and $h^4$. They require at every step respectively three and four computing functions, but despite this, there is a very accurate method.

If we compare the methods of Euler and Runge-Kutta, then the latter should be a significantly larger amount of calculations in one step of the algorithm. But this choice provides enhanced accuracy to $h^3$ or $h^4$. Ultimately, this enables to perform calculations with a large step. With its high accuracy and ease of implementation method Runge-Kutta 4-th order is one of the most common numerical methods for solving equations and systems of equations 1-th order.

When you select a step $h=0.1$ the accuracy of the solution method Runge-Kutta is achieved four significant digits. To achieve such accuracy in Euler method you need to choose a step $h=0.0001$ and $10^3$ to carry out calculations. Such a large number of computations necessarily lead to rounding errors.

The corresponding formulas for numerical methods of calculation are [11-15]:
1) Euler method:
$$y_{n+1} = y_n + hf(x_n, y_n), \tag{4}$$
where $y$ – the desired function, $f$ – the right part in the differential equation.
2) Runge-Kutta method of the second order:
$$y_{n+1} = y_n + \frac{h}{3}(k_1 + 2k_2), \tag{5}$$
$$k_1 = f(x_n, y_n),$$
$$k_2 = f(x_n + \frac{3}{4}h, y_n + \frac{3}{4}hk_1).$$
3) Runge-Kutta method of the third order:
$$y_{n+1} = y_n + \frac{h}{6}(k_1 + 4k_2 + k_3), \tag{6}$$
$$k_1 = f(x_n, y_n),$$
$$k_2 = f(x_n + \frac{h}{2}, y_n + \frac{h}{2}k_1),$$
$$k_3 = f(x_n + h, y_n - h(k_1 - 2k_2)).$$
4) Runge-Kutta method of the fourth order:
$$y_{n+1} = y_n + \frac{h}{6}(k_1 + 2k_2 + 2k_3 + k_4), \tag{7}$$
$$k_1 = f(x_n, y_n),$$
$$k_2 = f(x_n + \frac{h}{2}, y_n + \frac{h}{2}k_1),$$
$$k_3 = f(x_n + \frac{h}{2}, y_n + \frac{h}{2}k_2),$$
$$k_4 = f(x_n + h, y_n + hk_3).$$
5) Runge-Kutta method of the fifth order. The method to control the accuracy of the classroom integration is the method, for which at less than the number of calls to the right side can be complex formulas
$$y_{n+1} = y_n + \frac{h}{6}(k_1 + 4k_3 + k_4), \tag{8}$$

with the main ingredient error
$$y(x+h) - z(h) = r + O(h^6),$$
$$r = -\frac{1}{336}(42k_1 + 224k_3 + 21k_4 - 162k_5 - 125k_6),$$
$$k_1 = f(x_n, y_n),$$
$$k_2 = f(x_n + \frac{h}{2}, y_n + \frac{h}{2}k_1),$$
$$k_3 = f(x_n + \frac{h}{2}, y_n + \frac{h}{4}(k_1 + k_2)),$$
$$k_4 = f(x_n + h, y_n - h(k_2 - 2k_3)),$$
$$k_5 = f(x_n + \frac{2h}{3}, y_n + \frac{h}{27}(7k_1 + 10k_3 + k_4)),$$
$$k_6 = f(x_n + \frac{h}{5}, y_n + \frac{h}{625}(28k_1 - 125k_2 + 546k_3 + 54k_4 - 378k_5)).$$

If so $y(x+h) \approx z(h) + r$, then the method will receive a 5-th order accuracy.

## 5. Calculations of phase shifts

Spin states for a neutron-proton system are represented as $^{2S+1}L_J$, where $L$ - the moment of the system (the value of orbital moment $L=0; 1; 2; 3; 4;...$ correspond to S-, P-, D-, F-, G,... - states); $S$ - the spin of the system; $J$ - the complete moment of the system; $P=(-1)^L$ - parity. The spin states for pp- and nn-systems will be $^1S_0$-, $^3P_0$-, $^3P_1$-, $^1D_2$-, $^3F_3$- states. The spin states for np-systems will be $^1S_0$-, $^1P_1$-, $^3P_0$-, $^3P_1$-, $^1D_2$-, $^3D_2$- states.

By the phase-functions method it has been numerically obtained the phase shifts of np- scattering for Reid93 [6] and Argonne v18 [7] potentials. For solution phase equation (3) are selected numerical methods (4)-(8): Euler method (EM) and Runge-Kutta methods of the 2-, 3-, 4 - and 5th-order accuracy (RKM2, RKM3, RKM4 i RKM5). Program code for numerical calculations was written in the programming language FORTRAN.

The phase shifts were obtained with a precision of 0.001 degree for optimized selection steps ($h=0.1$) for numerical calculations. The phase shifts were at an output of the phase function $\delta_l(r)$ on an asymptotics at $r>10$ fm. The interval of energies was 1-350 MeV. The phase shifts are in degrees (Table 1). The masses of nucleons have been chosen as: $M_p=938.27231$ MeV; $M_n=939.56563$ MeV.

If we compare the phase shifts of np- scattering calculated for Reid93 and Argonne v18 potentials by different methods - on the basis of a solution of Schrödinger equation (see Ref. [6, 7]) and PFM (outcomes of the given paper), we may conclude that the discrepancy between the outcomes makes no more than seven percent. The best agreement with the data [6, 7] available for calculations performed in applying the Runge-Kutta methods of the 4-th and 5-th orders (the deviation is not more than two to four percent).

The comparison of the calculation results of the phase shifts for Reid93 and Argonne v18 potentials, obtained with the help of PFM, and the phase shifts for other potential models (Nijm I, Nijm II [6] i CD-Bonn [10]) and for the partial-wave analysis [6] indicates that the deviation between these data makes up to ten percents.

Along with the phase shifts in the problems of scattering one should deal with the scattering amplitudes, *S*- matrix elements and a number of other parameters. Based on the known phases of scattering one can obtain the complete amplitude, the full cross-section and the partial scattering amplitude accordingly [1]

$$F(\theta) = \frac{1}{k}\sum_{l=0}^{\infty}(2l+1)e^{i\delta_l}\sin\delta_l P_l(\cos\theta), \qquad (9)$$

$$\sigma = \frac{4\pi}{k^2}\sum_{l=0}^{\infty}(2l+1)\sin^2\delta_l, \qquad (10)$$

$$f_l = \frac{1}{k}e^{i\delta_l}\sin\delta_l, \qquad (11)$$

where $P_l(\cos\theta)$ - Legendre polynomials, $\theta$ - polar angle.

Depending on a computational method, the alteration of a scattering phase results in the respective alteration of the above-stated values $F(\theta)$, $\sigma$, $f_l$. For example, for $^3P_0$- state of the *np*- system an alteration of a scattering phase $\delta_l$ by the value $\varepsilon$ (1-2%) gives minor alteration (no more than 5%) of the real and imaginary part of partial amplitude $f_l$ (11).

In Ref. [16] of the values of the full cross-section of scattering $\sigma$, calculated by the phase shifts from Ref. [6, 7], and phase shifts according to the PFM (when solving the phase equations by the Runge-Kutta 4-th order) have been compared for NijmI, NijmII, Reid93 and Argonne v18 potentials. The difference between the calculations depending on the method of deriving scattering phases makes 0.1-6.3% for *np*- scattering. A similar comparison of the values $\sigma$, calculated *np* phase shifts from [6, 7] and according to PFM (for example, when solving the phase equations by the Runge-Kutta 4-th and 5-th orders) for potential Reid93 and Argonne v18, indicates that the difference between them is 4-5 percent. The results (in fm$^{-2}$) are shown in Table 2.

**Conclusions**

We have considered the known phase-functions method for the problem of a single-channel nucleon-nucleon scattering.

The expediency of application of Euler's method and Runge-Kutta method's for numerical calculations of phase shifts.

The phase-functions method has been used for the first time to calculate *np* phase shifts for the relevant spin configurations for nucleon-nucleon Reid93 and Argonne v18 potentials. Numerically obtained phase shifts well agree with the outcomes of original works [6, 7] for the same potentials (the deviation makes no more than seven percent). We have also compared the calculation results of the phase shifts using PFM with the phase shifts for other potential models (Nijm I, Nijm II i CD-Bonn) and for the partial-wave analysis: the deviation between these data makes up to ten percent.

In the following research we can apply these methods Runge-Kutta method to solve the coupled Schrödinger [17] to search the phase shifts of the mixed channels of the nucleon-nucleon scattering. In this case, the applied McHale-Thaler, Blatt-Biedenharn, Stapp or Matvienko-Ponomarev-Faifman parameterization's.

Table 1. *np* phase shifts for Reid93 and Argonne v18 potentials

| $T_{lab}$, MeV | Argonne v18 | | | | | Reid93 | | | | |
|---|---|---|---|---|---|---|---|---|---|---|
| | EM | RKM2 | RKM3 | RKM4 | RKM5 | EM | RKM2 | RKM3 | RKM4 | RKM5 |
| $^1S_0$ | | | | | | | | | | |
| 1 | 62.244 | 62.061 | 62.096 | 62.031 | 62.121 | 61.161 | 61.988 | 62.032 | 61.904 | 62.021 |
| 5 | 63.685 | 63.560 | 63.563 | 63.518 | 63.553 | 62.458 | 63.293 | 63.298 | 63.238 | 63.257 |
| 10 | 59.607 | 59.841 | 59.819 | 59.774 | 59.811 | 59.195 | 59.492 | 59.507 | 59.478 | 59.495 |
| 25 | 50.752 | 50.658 | 50.634 | 50.643 | 50.620 | 50.645 | 50.419 | 50.431 | 50.423 | 50.425 |
| 50 | 40.242 | 40.109 | 40.098 | 40.110 | 40.088 | 40.219 | 40.183 | 40.194 | 40.224 | 40.210 |
| 100 | 26.283 | 25.993 | 26.021 | 26.031 | 26.012 | 26.394 | 26.300 | 26.320 | 26.320 | 26.317 |
| 150 | 16.073 | 15.962 | 15.978 | 15.972 | 15.963 | 16.116 | 16.090 | 16.095 | 16.094 | 16.099 |
| 200 | 8.134 | 7.971 | 7.986 | 7.967 | 7.973 | 7.882 | 7.821 | 7.822 | 7.813 | 7.823 |
| 250 | 1.386 | 1.259 | 1.272 | 1.265 | 1.259 | 0.866 | 0.845 | 0.844 | 0.842 | 0.841 |
| 300 | -4.458 | -4.568 | -4.657 | -4.622 | -4.560 | -5.079 | -5.193 | -5.184 | -5.230 | -5.188 |
| 350 | -9.713 | -9.743 | -9.756 | -9.771 | -9.789 | -10.506 | -10.500 | -10.491 | -10.531 | -10.484 |
| $^3P_0$ | | | | | | | | | | |
| 1 | 0.182 | 0.182 | 0.197 | 0.192 | 0.195 | 0.172 | 0.172 | 0.173 | 0.181 | 0.173 |
| 5 | 1.678 | 1.676 | 1.667 | 1.654 | 1.675 | 1.607 | 1.602 | 1.621 | 1.622 | 1.622 |
| 10 | 3.791 | 3.785 | 3.792 | 3.787 | 3.791 | 3.655 | 3.643 | 3.654 | 3.663 | 3.643 |
| 25 | 8.442 | 8.435 | 8.442 | 8.432 | 8.423 | 8.273 | 8.245 | 8.247 | 8.261 | 8.247 |
| 50 | 11.131 | 11.120 | 11.124 | 11.146 | 11.148 | 11.142 | 11.123 | 11.126 | 11.149 | 11.146 |
| 100 | 8.807 | 8.798 | 8.801 | 8.813 | 8.840 | 9.297 | 9.208 | 9.212 | 9.235 | 9.211 |
| 150 | 3.870 | 3.865 | 3.867 | 3.865 | 3.862 | 4.505 | 4.443 | 4.445 | 4.464 | 4.456 |
| 200 | -1.381 | -1.359 | -1.356 | -1.371 | -1.368 | -0.732 | -0.768 | -0.764 | -0.754 | -0.765 |
| 250 | -6.337 | -6.353 | -6.351 | -6.362 | -6.355 | -5.881 | -5.832 | -5.836 | -5.843 | -5.833 |
| 300 | -11.025 | -11.009 | -11.006 | -11.090 | -11.062 | -10.596 | -10.611 | -10.606 | -10.624 | -10.607 |
| 350 | -15.347 | -15.318 | -15.325 | -15.334 | -15.326 | -15.055 | -15.068 | -15.061 | -15.072 | -15.062 |
| $^1P_1$ | | | | | | | | | | |
| 1 | -0.178 | -0.178 | -0.190 | -0.191 | -0.191 | -0.188 | -0.187 | -0.187 | -0.184 | -0.185 |
| 5 | -1.490 | -1.486 | -1.515 | -1.540 | -1.526 | -1.495 | -1.486 | -1.487 | -1.493 | -1.486 |
| 10 | -3.114 | -3.101 | -3.110 | -3.125 | -3.128 | -3.069 | -3.045 | -3.047 | -3.055 | -3.048 |
| 25 | -6.508 | -6.478 | -6.488 | -6.501 | -6.489 | -6.452 | -6.376 | -6.373 | -6.377 | -6.374 |
| 50 | -9.895 | -9.851 | -9.854 | -9.854 | -9.852 | -10.064 | -9.899 | -9.894 | -9.912 | -9.895 |
| 100 | -14.277 | -14.201 | -14.203 | -14.215 | -14.214 | -15.241 | -14.939 | -14.926 | -14.947 | -14.937 |
| 150 | -17.791 | -17.677 | -17.678 | -17.682 | -17.679 | -19.335 | -18.951 | -18.939 | -18.950 | -18.949 |
| 200 | -20.953 | -20.790 | -20.792 | -20.808 | -20.793 | -22.629 | -22.176 | -22.165 | -22.181 | -22.177 |
| 250 | -23.829 | -23.645 | -23.646 | -23.691 | -23.677 | -25.224 | -24.740 | -24.725 | -24.735 | -24.729 |
| 300 | -26.500 | -26.274 | -26.276 | -26.294 | -26.287 | -27.218 | -26.716 | -26.697 | -26.712 | -26.701 |
| 350 | -28.969 | -28.706 | -28.704 | -28.725 | -28.716 | -28.777 | -28.185 | -28.161 | -28.165 | -28.167 |
| $^3P_1$ | | | | | | | | | | |
| 1 | -0.104 | -0.104 | -0.105 | -0.112 | -0.107 | -0.104 | -0.103 | -0.107 | -0.110 | -0.104 |
| 5 | -0.924 | -0.922 | -0.923 | -0.930 | -0.925 | -0.924 | -0.921 | -0.931 | -0.940 | -0.932 |
| 10 | -2.052 | -2.046 | -2.047 | -2.052 | -2.046 | -2.054 | -2.047 | -2.048 | -2.053 | -2.047 |
| 25 | -4.850 | -4.829 | -4.830 | -4.843 | -4.829 | -4.879 | -4.854 | -4.856 | -4.864 | -4.858 |
| 50 | -8.211 | -8.161 | -8.162 | -8.181 | -8.168 | -8.327 | -8.268 | -8.269 | -8.271 | -8.269 |
| 100 | -13.191 | -13.087 | -13.088 | -13.104 | -13.089 | -13.504 | -13.387 | -13.388 | -13.402 | -13.390 |
| 150 | -17.448 | -17.311 | -17.313 | -17.330 | -17.315 | -17.849 | -17.695 | -17.696 | -17.715 | -17.706 |
| 200 | -21.404 | -21.254 | -21.255 | -21.275 | -21.257 | -21.751 | -21.568 | -21.564 | -21.568 | -21.568 |
| 250 | -25.158 | -24.989 | -24.991 | -25.018 | -24.992 | -25.311 | -25.097 | -25.098 | -25.142 | -25.109 |
| 300 | -28.721 | -28.534 | -28.536 | -28.550 | -28.538 | -28.564 | -28.329 | -28.330 | -28.353 | -28.331 |
| 350 | -32.099 | -31.894 | -31.896 | -31.895 | -31.895 | -31.544 | -31.298 | -31.300 | -31.311 | -31.305 |

| | | | | | ¹D₂ | | | | | |
|---|---|---|---|---|---|---|---|---|---|---|
| 1 | 0.001 | 0.001 | 0.001 | 0.000 | 0.000 | 0.001 | 0.001 | 0.001 | 0.000 | 0.000 |
| 5 | 0.037 | 0.037 | 0.042 | 0.041 | 0.041 | 0.037 | 0.038 | 0.037 | 0.040 | 0.038 |
| 10 | 0.151 | 0.152 | 0.157 | 0.164 | 0.156 | 0.150 | 0.152 | 0.154 | 0.167 | 0.155 |
| 25 | 0.679 | 0.680 | 0.681 | 0.691 | 0.680 | 0.672 | 0.673 | 0.673 | 0.670 | 0.671 |
| 50 | 1.703 | 1.705 | 1.706 | 1.722 | 1.715 | 1.689 | 1.692 | 1.693 | 1.702 | 1.693 |
| 100 | 3.807 | 3.813 | 3.814 | 3.825 | 3.817 | 3.784 | 3.797 | 3.798 | 3.824 | 3.807 |
| 150 | 5.720 | 5.728 | 5.730 | 5.746 | 5.739 | 5.663 | 5.688 | 5.690 | 5.690 | 5.691 |
| 200 | 7.307 | 7.311 | 7.314 | 7.324 | 7.323 | 7.225 | 7.264 | 7.266 | 7.283 | 7.267 |
| 250 | 8.566 | 8.542 | 8.545 | 8.553 | 8.555 | 8.505 | 8.556 | 8.559 | 8.572 | 8.560 |
| 300 | 9.511 | 9.450 | 9.453 | 9.460 | 9.454 | 9.568 | 9.630 | 9.635 | 9.651 | 9.637 |
| 350 | 10.163 | 10.080 | 10.085 | 10.122 | 10.088 | 10.477 | 10.551 | 10.555 | 10.545 | 10.552 |
| | | | | | ³D₂ | | | | | |
| 1 | 0.004 | 0.006 | 0.008 | 0.010 | 0.008 | 0.006 | 0.007 | 0.008 | 0.011 | 0.009 |
| 5 | 0.199 | 0.201 | 0.221 | 0.222 | 0.222 | 0.223 | 0.223 | 0.223 | 0.223 | 0.223 |
| 10 | 0.816 | 0.818 | 0.847 | 0.850 | 0.848 | 0.852 | 0.852 | 0.851 | 0.852 | 0.852 |
| 25 | 3.695 | 3.708 | 3.713 | 3.721 | 3.722 | 3.736 | 3.738 | 3.744 | 3.751 | 3.739 |
| 50 | 8.882 | 8.940 | 8.946 | 8.964 | 8.958 | 9.010 | 9.006 | 9.012 | 9.052 | 9.051 |
| 100 | 16.947 | 17.101 | 17.109 | 17.123 | 17.118 | 17.078 | 17.104 | 17.140 | 17.154 | 17.155 |
| 150 | 21.783 | 21.839 | 21.851 | 21.875 | 21.862 | 21.713 | 21.679 | 21.726 | 21.740 | 21.731 |
| 200 | 24.154 | 24.180 | 24.194 | 24.227 | 24.194 | 23.862 | 23.903 | 23.960 | 23.992 | 23.968 |
| 250 | 25.009 | 25.034 | 25.048 | 25.096 | 25.144 | 24.738 | 24.737 | 24.814 | 24.825 | 24.810 |
| 300 | 25.044 | 24.981 | 24.996 | 25.052 | 25.049 | 24.765 | 24.764 | 24.835 | 24.843 | 24.842 |
| 350 | 24.460 | 24.375 | 24.389 | 24.413 | 24.395 | 24.379 | 24.319 | 24.374 | 24.421 | 24.397 |

Table 2. The full cross-section of *np*- scattering

| | Argonne v18 | | | | Reid93 | | | |
|---|---|---|---|---|---|---|---|---|
| $T_{lab}$, MeV | EM | RKM3 | RKM5 | [7] | EM | RKM3 | RKM5 | [6] |
| 1 | 204.08 | 203.52 | 203.62 | 203.23 | 199.96 | 203.28 | 203.24 | 202.74 |
| 5 | 210.78 | 210.35 | 210.33 | 209.47 | 206.23 | 209.34 | 209.19 | 209.07 |
| 10 | 200.87 | 201.73 | 201.72 | 201.42 | 198.95 | 200.16 | 200.09 | 199.94 |
| 25 | 194.34 | 193.76 | 193.65 | 193.12 | 193.21 | 191.89 | 191.85 | 191.70 |
| 50 | 209.12 | 208.51 | 208.70 | 207.67 | 211.19 | 209.99 | 210.45 | 209.81 |
| 100 | 274.10 | 274.12 | 274.45 | 273.40 | 286.58 | 283.90 | 284.19 | 283.50 |
| 150 | 359.15 | 358.08 | 358.27 | 357.63 | 375.04 | 370.62 | 370.90 | 370.21 |
| 200 | 448.97 | 446.64 | 446.71 | 446.39 | 461.92 | 457.53 | 457.83 | 457.23 |
| 250 | 540.30 | 537.27 | 539.34 | 537.10 | 549.77 | 543.86 | 543.95 | 543.29 |
| 300 | 635.35 | 629.57 | 630.86 | 629.36 | 635.33 | 628.76 | 628.96 | 628.32 |
| 350 | 730.21 | 722.75 | 723.06 | 722.59 | 721.57 | 712.13 | 712.62 | 711.73 |